\documentclass[a4paper,10pt,notitlepage]{article}
\usepackage{amsfonts, amssymb, amsmath, latexsym}
\usepackage{color}

\def\g{\gamma}

\def\RM{{\mathbf R}}

\def\sqd{\frac{\nu_1}{\sigma_1}}

\title{Counterparty risk valuation for CDS.}
\author{Christophette Blanchet-Scalliet\\
Universit\'e de Lyon\\
CNRS, UMR 5208, Institut Camille Jordan\\
Ecole Centrale de Lyon\\
Universit\'e Lyon 1\\
INSA de Lyon\\
36 avenue Guy de Collongue\\
69134 Ecully Cedex - FRANCE\\
\\
Fr\'ed\'eric Patras\footnote{Consultant, Zeliade Systems.}\\
Universit\'e de Nice et CNRS\\
UMR 6621, J.-A. Dieudonn\'e\\
Parc Valrose\\
06108 Nice Cedex 2 - FRANCE\\}
\date{Version of \today}

\begin{document}
\maketitle
%\tableofcontents
\newtheorem{lem}{Lemma}[section]
\newtheorem{defi}[lem]{Definition}
\newtheorem{prop}[lem]{Proposition}
\newtheorem{cor}[lem]{Corollary}
\newtheorem{theo}[lem]{Theorem}

%%%%%%%%%%%%%%%%%%%%%%%%%%%%%%%%%%%%%%%%%%%%%%%%%%%%%%%%%%%%%
\begin{abstract}
The valuation of counterparty risk for single name credit derivatives requires the computa-
tion of joint distributions of default times of two default-prone entities.
For a Merton-type model, we derive some formulas for these joint distribu-
tions. As an application, closed formulas for counterparty risk on a CDS
or for a first-to-default swap on two underlyings are obtained.
\end{abstract}

%%%%%%%%%%%%%%%%%%%%%%%%%%%%%%%%%%%%%%%%%%%%%%%%%%%%%%%%%%%%%%%%%%%%%%%%%%%%%%%%%%%%%%%%%%%%%%%%%%%%%%%%%%%%%%%%%%%%%%%%
\section{Introduction}
Default Risk has become one of the key issues of contemporary finance. In practice, default risk computations are directly involved in three, closely connected, but still separated areas: first, the pricing of single and multiname credit derivatives (ABS, CDS, CDOs...), second, risk valuation for regulatory, risk management and economic capital valuation purposes, third, rating assignements. The present article is focused mainly on the first two aspects of risk valuation. Notice however that the next generation of rating methodologies will most probably have to include advanced correlation tools such as the ones considered here. 

Although deriving new formulas for credit derivative transactions on two assets, our main interest will be on counterparty risk. Once again, this is a central argument, both for regulatory purposes (counterparty risk was already a key issue in the Basel 1 agreements, but its relevance and complexity was emphasized by the later improvements of the set of rules, see e.g. \cite{BIS1,hull,BCBS}) and also for practical risk management. 

Concretely, the 2007 ``subprime'' credit crisis has emphasized how a mispricing of counterparty risk could be dangerous. Before the crisis, financial institutions engaged heavily in credit derivatives transactions for hedging purposes. One of the main side effects of the crisis has been to reveal that their counterparties were not reliable: monoline insurers (who act traditionaly as guarantees for municipalities-type bond emissions but engaged more recently in the ABS business) faced downgrades and/or bankruptcy, whereas some companies offering CDS protection have also been on the wedge of bankruptcy due to spread widenings and collateral agreements.
In spite of these facts, in the present state of the art quantitative counterparty risk assessment is still largely in infancy. 

We address here one of the fundamental questions in the field, namely we introduce a Merton-type model and derive formulas for the joint distributions of the default events -and other relevant joint probabilities- of two default-prone entities. Closed formulas for counterparty risk on a CDS transaction or for a first-to-default swap on two underlyings follow. 

Notice that we refrain to seek for the outmost generality or for a complete list of pricing formulas that could be obtained using the methods developed in the present article. Most credit derivatives on two underlyings with time-dependent payoffs can actually be priced using our results. The exercise of deriving the corresponding formulas is left to the interested reader. 

%%%%%%%%%%%%%%%%%%%%%%%%%%%%%%%%%%%%%%%%%%%%%%%%%%%%%%%%%%%%%%%%%%%%%%%%%%%%%%%%%%%%%%%%%%%%%%%
\section{Modeling two dimensional default risk.}
%%%%%%%%%%%%%%%%%%%%%%%%%%%%%%%%%%%%%%%%%%%%%%%%%%%%%%%%%%%%%%%%%%%%%%%%%%%%%%%%%%%%%%%%%%%%%%%
Recall that we are interested in modeling the counterparty risk on a credit risk transaction. This involves (in general and depending of the particular features of the transaction) the computation of various joint laws. For example, for a vanilla credit default swap, the computation relies on the distribution of the default time of the counterparty and the conditional distribution (w.r. to the default of the counterparty) of the values of the CDS contract. The same methodology can be applied to the valuation of first-to-default contracts on two underlyings when the payoff is time-dependent (see section 4 of the present article; the time-independent case has been addressed in \cite{hkr,z,p}).

We follow Merton's structural approach to risk, where the default of a company is triggered by its firm value falling below a threshold (see \cite{me} or e.g. \cite[Chap. 3]{br}). It is well-known that the original Merton methodology, where the threshold is determined by the firm's liabilities, as well as its natural refinements such as the Black-Cox first passage time formulas \cite{bc}, give qualitatively good but numerically poor results. For example, the structural method will rank correctly the risks of two companies but will not be able to predict the spreads of the corresponding CDS. 
For pricing purposes, the correct way to use the Merton and Black-Cox models is to calibrate their parameters on the market prices of the securities issued by the companies. Doing so insures that the models price correctly the risk. This is the ground for their various uses, either in credit risk measurement (think to the Basel 2 Vasicek-type large pool formulas \cite{vas,hull}), either for pricing purposes (think to the one-factor Gaussian copula model for CDO tranches \cite{s}).  

Notice that the computations we are interested in require to go beyond the existing formulas for two-dimensional credit risk in the structural approach \cite{hkr,z,p}. Indeed, the formulas in these articles rely mainly on the computation of the probability that one (and only one) out of two firms defaults or that the two firms default before a given maturity, whereas pricing counterparty risk for a CDS transaction requires (among others) the exact knowledge of default times distributions.

We have tried to keep some balance between generality and the choice of too restrictive assumptions. Notice that the model may be simplified by selecting a suitable subset of meaningful parameters, depending on the particular situations and/or the available set of data. In the end, the model we use is as follows. 

We label 1 and 2 the two default-prone entities. In view of applications to counterparty risk, the counterparty will be labelled 2. 
We write $\tau_1$ (resp. $\tau_2$) for the random time when the first (resp. second) entity defaults.
Since we work in the structural model, defaults are triggered when two log-normal processes $V_1$ and $V_2$ associated respectively to the first and second entity fall below a barrier process. As in Black-Cox \cite{bc}, we assume  that the barrier is described by a time-dependent deterministic process $v_i(t):=K_ie^{\gamma_i t}$. At last, we assume that the risk-neutral dynamics of the processes $V_1$ and $V_2$ are given by:
$$\frac{dV_i(t)}{ V_i(t)}=(r-k_i) dt+\sigma_idB_i(t),$$
where $r$ is the constant short-term interest rate and $B_i,\ i=1,2$ are two Brownian motions.
The coefficient $k_i$ is a payout ratio representing net payouts/inflows by the firm, see e.g. the account of Merton's approach in \cite[Chap.2]{br}. 
Notice that we use a risk-neutral dynamics since we are interested in pricing formulas, but the same construction holds with a real-life dynamics (just replace $r-k_i$ by the drift under the historical probability). 

At last, we assume that the Brownian motions $B_1$ and $B_2$ are correlated: $Cov (B_1(t),B_2(t))=\varrho t$. This is an important assumption that reflects the fact that the counterparty of a CDS transaction is usually a well-rated company that will not default excepted in a very bad macroeconomic environment. In particular, if the counterparty defaults, one may expect CDS spread on other entities to widen considerably simultaneously. This phenomenon is accounted for by a positive correlation coefficient. It is the main reason, together with the dynamical aspects of the problem, why simplistic counterparty risk models fail, both theoretically and empirically, to account for counterparty-driven loss expectations on derivative contracts.

%%%%%%%%%%%%%%%%%%%%%%%%%%%%%%%%%%%%%%%%%%%%%%%%%%%%%%%%%%
\section{Counterparty risk}
%%%%%%%%%%%%%%%%%%%%%%%%%%%%%%%%%%%%%%%%%%%%%%%%%%%%%%%%%%%%%%

In this section, we derive a closed formula for the counterparty default leg of a CDS contract, that is the present value (PV) of the expected losses on a vanilla single-name credit derivative transaction due to the default of the protection seller.   
Notice that the  notion of counterparty default leg should not be confused with the usual notion of default leg for a CDS contract, that is, the present value of future payments by the protection seller.

We follow standard market practices. Namely, we assume that the CDS contract has a notional value $C$ and that, if the underlying entity defaults, the buyer of protection receives $(1-R_u)\cdot C$ from the seller of protection, where $R_u$ stands for the recovery rate of the underlying entity. Similarly, if the counterparty (the protection seller) defaults at $t$ on the CDS contract, we assume that the buyer of protection receives $(1-R_c)\cdot V_t^+$, where $V_t^+=\sup (0, V_t)$ stands for the positive part of the market value of the CDS contract at $t$, and where $R_c$ stands for the counterparty recovery rate. The present moment is normalized to $t=0$ in what follows, it may or may not coincide with the inception of the CDS contract.

%%%%%%%%%%%%%%%%%%%%%%%%%%%%%%%%%%%%%%%%%%%%%%%%%%%%%%%%%%%%%%%
\subsection{Counterparty default leg}
%%%%%%%%%%%%%%%%%%%%%%%%%%%%%%%%%%%%%%%%%%%%%%%%%%%%%%%%%%%%%%%%

Using the joint law of the default time of the counterparty and of the underlying entity for the CDS, we compute the default probability of the underlying entity conditional to the default of the counterparty, and obtain a closed formula for the counterparty default leg. 

The coupon payments rate (CDS spread) by the protection buyer is written $s$. We assume continuous payments of the fees: in practice, fees are payed quarterly (and sometimes semi-annually), but due to the payments-in-arrears conventions, the continuous payments assumption is a reasonable one. It is usually defined, at inception of the contract, by equalizing the fee and default legs of the CDS contract (see Section~\ref{3.3}. Some empirical adjustment is sometimes done for the counterparty default risk, taking into account advanced risk management parameters such as netting agreements. Our computation of the counterparty default leg leads to a new, sounder, methodology, to assess the fair value of this adjustment.

First, the theoretical counterparty default leg $D_c$ is given by
\begin{eqnarray*}
D_c&=& (1-R_c)\cdot E[e^{-r\tau_2}\cdot \sup (0,p(V_1(\tau_2),\tau_2))1_{\tau_2<(T\wedge\tau_1)}]
\end{eqnarray*}
where $p(V_1(\tau_2),\tau_2)$ is the market price of the CDS contract at $t=\tau_2$ (when $\tau_1\geq \tau_2$). The value $p(V_1(\tau_2),\tau_2)$ is obtained as the difference between the default and fee legs, that is:
$$p(V_1(\tau_2),\tau_2)=D_l(V_1(\tau_2),\tau_2)- \frac{sC}{r}\cdot E[(1-e^{-r((T\wedge\tau_1)-\tau_2)})1_{\tau_1\geq \tau_2}|{\cal F}_{\tau_2}]$$
where we write ${\cal F}_t$, as usual, for the natural filtration of the probability space underlying the two Brownian motions $B_1(t), B_2(t)$. 

Notice that the term $T\wedge\tau_1$ can be, in most situations, safely replaced by $T$ in the previous expansion. This is because, for standard values for the spreads and implied default probabilities in single-name default risk computations, the computation of the fee leg conditional to the hypothesis that no default occurs is a good first-order approximation to the unconditional fee leg. The following computations could be simplified accordingly.
 
\vspace{0.5cm}
\noindent Now, the value of  $D_l(V_1(\tau_2),\tau_2)$ is given by:

\begin{eqnarray*}
D_l(V_1(\tau_2),\tau_2)&=&E[C(1-R_u)e^{-r(\tau_1-\tau_2)}1_{\tau_2\leq\tau_1 \leq T}|\mathcal {F} _{\tau_2}]\\
\end{eqnarray*}

\begin{theo}
\def\ssqu{\alpha}
\def\ssqd{\beta}
The counterparty default leg $D_c$ is given by:
$$D_c= C(1-R_c)$$
$$*E\Bigg[1_{\tau_2<(T\wedge\tau_1)}\Bigg(e^{-r\tau_2}(1-R_u+\frac{s}{r})\bigg(e^{-\mu_{\tau_2}(\ssqd -\ssqu )}N(\frac{-\mu_{\tau_2}-\ssqu (T-\tau_2)}{\sqrt{T-\tau_2}})
+e^{-\mu_{\tau_2}(\ssqd +\ssqu )} N(\frac{-\mu_{\tau_2}+\ssqu (T-\tau_2)}{\sqrt{T-\tau_2}})\bigg)$$
$$
 -\frac{s}{r}\bigg(1-e^{-r(T-\tau_2)}\Big(1-N(\frac{-\mu_{\tau_2}-\beta (T-\tau_2)}{\sqrt{T-\tau_2}})-e^{-2\mu_{\tau_2}\beta}N(\frac{-\mu_{\tau_2}+\beta (T-\tau_2)}{\sqrt{T-\tau_2}})\Big)\bigg)\Bigg)_{+}\Bigg]$$
where $\nu_1:=r-k_1-\gamma_1-\frac{1}{2}\sigma^2_1$, $\alpha :=\sqrt{\frac{\nu_1^2}{\sigma_1^2}+2r}$, $\beta:=\sqd$, $\mu_{\tau_2}:=\frac{\ln (\frac{V_1(\tau_2)}{v_1(\tau_2)})}{\sigma_1}$.
\end{theo}

Indeed,
$$p(V_1(\tau_2),\tau_2)1_{\tau_2<(T\wedge\tau_1)}=C1_{\tau_2<(T\wedge\tau_1)}E\Big[(1-R_u)e^{-r(\tau_1-\tau_2)}1_{\tau_2 \leq \tau_1 <T}-\frac{s}{r}1_{\tau_2 <\tau_1}$$
$$
+\frac{s}{r}e^{-r(\tau_1-\tau_2)}1_{\tau_2 \leq \tau_1 <T}+\frac{s}{r}e^{-r(T-\tau_2)}1_{\tau_2 \leq T <\tau_1} |\mathcal {F} _{\tau_2}\Big]$$
and therefore
$$p(V_1(\tau_2),\tau_2)1_{\tau_2<(T\wedge\tau_1)}=C1_{\tau_2<(T\wedge\tau_1)}\Big(E\Big[(1-R_u+\frac{s}{r})e^{-r(\tau_1-\tau_2)}1_{\tau_2 \leq \tau_1 <T}|\mathcal {F} _{\tau_2}\Big]
-E\Big[\frac{s}{r}1_{\tau_2 <\tau_1}|\mathcal {F} _{\tau_2}\Big]$$
$$
+E\Big[\frac{s}{r}e^{-r(T-\tau_2)}1_{\tau_2 \leq T <\tau_1} |\mathcal {F} _{\tau_2}\Big]\Big).
$$

The Theorem follows by standard computations of integrals over Gaussian densities that are sketched in the Appendix 1.

\subsection{Explicit formulas}

Recall that $\tau_i ={\rm inf}\{t, V_i(t)\leq K_ie^{\g_it}\}$.
The condition $V_i(t)\leq K_ie^{\g_it}$ can be rewritten: ${W}_i(t)\geq y_0^i,$
where ${W}_i(t)={\rm ln}({K_ie^{\g_it}\over V_i(t)})-{\rm ln}({K_i\over V_i(0)})$ and $y_0^i={\rm ln}\ V_i(0)-{\rm ln}\ K_i$. Equivalently,
${W}_i(t)$ is the diffusion process:
$$d{W}_i(t)=-\nu_i t-\sigma_idB_i(t),$$
with ${\bf W}_i(0)=0$ and $\nu_i:=r-k_i-\gamma_i-\frac{1}{2}\sigma_i^2$.\par
Let us define ${\bf Z }(t)$ by:
$${\bf Z }(t)=(Z_1(t),Z_2(t))^\ast ={1\over {\sqrt{1-\varrho^2}}}\left({\begin{array}{cc}
\sigma_1^{-1}&-\varrho\sigma_2^{-1}\\
0&{\sqrt{1-\varrho^2}}\sigma_2^{-1}\end{array}}\right)\left({\begin{array}{c}y_0^1-W_1(t)\\
y_0^2-W_2(t)\end{array}}\right).$$
We get:
$$dZ_1(t)=\phi_1 dt+dX_1(t),\ dZ_2(t)=\phi_2 dt+dX_2(t),$$
where ${\bf X}(t)$ is a standard planar Brownian motion and $$\phi_1={{\nu_1\sigma_2}-{\nu_2\sigma_1}\varrho \over {\sigma_1\sigma_2\sqrt{1-\varrho^2}}},
\phi_2={{\nu_2}\over \sigma_2}.$$
In particular, ${\bf Z}(t)$ is a 2-dim. Brownian motion with drift and the barrier conditions $V_i(t)=v_i(t)$ now read:
$Z_2(t)= 0$ and
$\ {\sqrt{1-\varrho^2}}Z_1(t)+\varrho Z_2(t)= 0.$

\noindent We want to compute the default leg $D_c$, that is, equivalently:
\begin{eqnarray*}
	E[h(B_1(\tau_2),\tau_2)1_{\tau_2\leq (T\wedge\tau_1)}]&=&E[\tilde{h}(Z_1(\tau_2),\tau_2)1_{\tau_2\leq (T\wedge \tau_1)}]\\
	&=&\int_0^T\int_0^{+\infty}\tilde{h}(r,s)P(\tau_2 \in ds,Z_1(\tau_2) \in da) ds da
	\label{eq:}
\end{eqnarray*}
where 
$h(x,t):= $
$$\Bigg(e^{-rt}(1-R_u+\frac{s}{r})\bigg(e^{-\mu_{x,t}(\beta -\alpha )}N(\frac{-\mu_{x,t}-\alpha (T-t)}{\sqrt{T-t}})
+e^{-\mu_{x,t}(\beta +\alpha )} N(\frac{-\mu_{x,t}+\alpha (T-t)}{\sqrt{T-t}})\bigg)$$
$$
 -\frac{s}{r}\bigg(1-e^{-r(T-t)}\Big(1-N(\frac{-\mu_{x,t}-\beta (T-t)}{\sqrt{T-t}})-e^{-2\mu_{x,t}\beta}N(\frac{-\mu_{x,t}+\beta (T-t)}{\sqrt{T-t}})\Big)\bigg)\Bigg)_{+},$$
$\mu_{x,t}:=\sigma_1^{-1}(\nu_1t+\sigma_1x+\ln V_1(0)-\ln K_1)$ and
$\tilde{h}(z,t)=h(\frac{-y^1_0-\nu_1t}{\sigma_1}+\sqrt{1-\rho^2} \cdot z,t)$.

\noindent Applying the Girsanov theorem, $(Z_1(t),Z_2(t))^\ast$ is a classical Brownian motion for the probability law $\bf Q$:
$${d{\bf Q}\over d{\bf P}}=e^{-\phi_1X_1(T)-\phi_2X_2(T)-[{\phi_1^2\over 2}+{\phi_2^2\over 2}]T}\ \ {\bf P}a.s.$$

\noindent Let $r_0 e^{i\theta_0}:=Z_1(0)+i Z_2(0)={y_0^1\sigma_2 -\varrho y_0^2\sigma_1\over \sigma_1\sigma_2\sqrt{1-\varrho^2}}+i\ {y_0^2\over\sigma_2}$.

\begin{lem}\label{prob}
We have, for $(a,0)\in\RM^2$ s.t. $a> 0$ :
\begin{eqnarray*}
{\bf P}(\tau_2 \in dt, \tau_2=\tau_2\wedge\tau_1, Z_1({\tau_2}) \in da)&=& e^{\phi_1(a-r_0\cos (\theta_0))-\phi_2r_0\sin (\theta_0)-{||\vec{\phi}||^2t\over 2}}\frac{\pi}{\alpha^2 t a}e^{-(a^2+r_0^2)/2t}\\
&&\sum_{n=0}^\infty  n \sin \frac{n\pi \theta_0}{\alpha}I_{n\pi/\alpha}(\frac{ar_0}{t})dadt.
\end{eqnarray*}
where $||\vec{\phi}||^2:={\phi_1^2+\phi_2^2}$, $\alpha:=\arcsin(\varrho)+\frac{\pi}{2}$ and $I_{n\pi/\alpha}$ is the modified Bessel function of index ${n\pi/\alpha}$.

\end{lem}

The Lemma follows from Thm.~\label{app1} in Appendix 2 (up to a straightforward adaptation since we consider here a polyhedral domain $D:=\{ (x,y)\in\RM^2|y\geq 0,\sqrt{1-\varrho^2}x+\varrho y\geq 0\}$ with a horizontal lower --instead of an upper-- boundary, so that the signs have to be changed accordingly in the formulas). 

Indeed, according to \cite{cj} (see also \cite[pp.697]{p}), we have, for $(a,b)$ in a neighborhood of $(a,0)$ with $b>0$:
 $$f(a,b,t)dadb={\bf Q}({\bf Z}(t)\in (da,db),\tau_1\wedge\tau_2>t)$$
 $$=\frac{2\mu}{\alpha t}e^{-(\mu^2+r_0^2)/2t}\sum_{n=0}^\infty  \sin \frac{n\pi \theta}{\alpha}\sin \frac{n\pi \theta_0}{\alpha}I_{n\pi/\alpha}(\frac{\mu r_0}{t})d\mu d\theta ,$$
 where $\theta :=arctg (\frac{b}{a} ),\ \mu:=\sqrt{a^2+b^2}$.
 Therefore, we have:
\begin{eqnarray*}
{\bf Q}(\tau_2=\tau_2\wedge\tau_1,\tau_2 \in dt, Z_1({\tau_2}) \in da)&=&\frac{\pi}{\alpha^2 t a}e^{-(a^2+r_0^2)/2t}\sum_{n=0}^\infty  n \sin \frac{n\pi \theta_0}{\alpha}I_{n\pi/\alpha}(\frac{ar_0}{t})dadt,
\end{eqnarray*}
 and the proof follows.

\begin{theo}\label{counter}
The counterparty default leg $D_c$ of the CDS is given by:

\begin{eqnarray*}
D_c&=&C(1-R_c)\int_0^T\int_0^{+\infty}\tilde{h}(\mu ,t)e^{\phi_1(\mu -r_0\cos (\theta_0))-\phi_2r_0\sin (\theta_0)-{||\vec{\phi}||^2t\over 2}}\\
&&\frac{\pi}{\alpha^2 t \mu }e^{-(\mu ^2+r_0^2)/2t}\sum_{n=0}^\infty  n \sin \frac{n\pi \theta_0}{\alpha}I_{n\pi/\alpha}(\frac{\mu r_0}{t})dtd\mu .{\nonumber}\\
\end{eqnarray*}

\end{theo}

Notice that although involving a double integral on Gaussian densities and Bessel function (which asymptotic behavior is well-understood, making numerical approximations easy and efficient), this kind of higher order integrals is highly familiar in physics (heat conduction in solids, cross-sections computations, acoustics...) and can be handled efficiently. However these techniques are not, for the time being, of our competence, and we postpone to further work the numerical analysis of the problem. For the use of Bessel functions in physics, we refer to the classical treatises such as \cite{cj,gm,w}.

%%%%%%%%%%%%%%%%%%%%%%%%%%%%%%%%%%%%%%%%%%%%%%%%%%%%%%%%%%%%%%%%%%%%%%%%%%%%%%%%%
\subsection{Fair price of a CDS}\label{3.3}
%%%%%%%%%%%%%%%%%%%%%%%%%%%%%%%%%%%%%%%%%%%%%%%%%%%%%%%%%%%%%%%%%%%%%%%%%%%%%%%%%

In this section, we take advantage of our computation of the counterparty default leg to compute the fair price of a CDS contract at an arbitrary time (normalized again to $t=0$ in this section) between the inception of the contract and its maturity $T$. As usual, it is obtained as difference between the default and fee legs, but, contrary to the usual pricing formulas, we take into account exactly the effect of the counterparty default leg.

Notice that we insist in deriving exact formulas but, for practical purposes, it may be convenient to use the standard approximation schemes to simplify the use of the formulas and fasten the computations. For example, in order to compute the fee leg, it may be convenient to assume that defaults can occur only at times $t_i, i=1...n$, so that the following integral expressions can be safely replaced by finite sums involving only the computation of the default probabilities ${\bf P}(\tau_1\wedge\tau_2\leq t)$.

Let us write 
$D_s$ for the ``standard'' default leg of the CDS, that is, the present value of the cash-flows corresponding to payments by the seller of protection if there is a default occurring on the assets underlying the CDS contract. The total default leg $D_{tot}$ of the CDS contract is given by the sum $D_s+D_c$, where $D_c$ is given by Thm.~\ref{counter}.

\begin{lem}
The standard default leg is given by:
$$D_s=C(1-R_u)\int_0^T\int_0^{+\infty}e^{\phi_1(\mu \cos(\alpha)-r_0\cos(\theta_0))+\phi_2(\mu \sin(\alpha)-r_0\sin(\theta_0))-{||\vec{\phi}||^2T\over 2}}\frac{\pi}{\alpha^2 t \mu }$$
$$e^{-(\mu ^2+r_0^2)/2t}\sum_{n=0}^\infty (-1)^{n+1} n \sin \frac{n\pi \theta_0}{\alpha}I_{n\pi/\alpha}(\frac{\mu r_0}{t})d\mu dt$$
\end{lem}

The proof follows from the same reasoning as the one of Lemma~\ref{prob} and can be omitted.

Notice that:
$${\bf Q}(\tau_1\in dt,\tau_1=\tau_1\wedge\tau_2, Z_1(t)+iZ_2(t)=d\mu \cdot e^{i\alpha})=$$
$$\frac{\pi}{\alpha^2 t \mu }e^{-(\mu ^2+r_0^2)/2t}\sum_{n=0}^\infty (-1)^{n+1} n \sin \frac{n\pi \theta_0}{\alpha}I_{n\pi/\alpha}(\frac{\mu r_0}{t})d\mu dt$$

Let us write $s$ for the spread of the CDS contract (the continuous coupon rate served by the protection buyer to the protection seller, till the end of the contract, or the default of the underlying, or the default of the protection seller); $s$ is fixed at inception of the contract and remains constant till the maturity $T$. Recall that the fee leg of a credit derivative contract is the present value of cumulated payments by the protection buyer. We write from now on $\tau$ for $\tau_1\wedge\tau_2$.

\begin{lem}\label{fee}
The fee leg $F$ of the CDS contract is given by:
$$Fs^{-1}=\frac{1-e^{-rT}}{r} {\bf P}(\tau \geq T)+E[\frac{1-e^{-r\tau}}{r}1_{\tau\leq T}]$$
$$=\frac{1-e^{-rT}}{r} \int_0^\infty\int_0^\alpha e^{\phi_1(\mu \cos(\kappa)-r_0\cos(\theta_0))+\phi_2(\mu \sin(\kappa)-r_0\sin(\theta_0))-{||\vec{\phi}||^2T\over 2}}\frac{2\mu }{\alpha T}$$
$$e^{-(\mu ^2+r_0^2)/2T}\sum_{n=1}^\infty \sin \frac{n\pi\kappa}{\alpha}\sin \frac{n\pi \theta_0}{\alpha}I_{n\pi/\alpha}(\frac{\mu r_0}{T})d\mu d\kappa$$
$$+\int_0^T\int_0^{+\infty}\frac{1-e^{-rt}}{r}e^{\phi_1(\mu \cos(\alpha)-r_0\cos(\theta_0))+\phi_2(\mu \sin(\alpha)-r_0\sin(\theta_0))-{||\vec{\phi}||^2T\over 2}}\frac{\pi}{\alpha^2 t \mu }$$
$$e^{-(\mu ^2+r_0^2)/2t}\sum_{n=0}^\infty (-1)^{n+1} n \sin \frac{n\pi \theta_0}{\alpha}I_{n\pi/\alpha}(\frac{\mu r_0}{t})d\mu dt$$
$$+\int_0^T\int_0^{+\infty}\frac{1-e^{-rt}}{r}e^{\phi_1(\mu -r_0\cos (\theta_0))-\phi_2r_0\sin (\theta_0)-{||\vec{\phi}||^2t\over 2}}\\
\frac{\pi}{\alpha^2 t \mu }e^{-(\mu ^2+r_0^2)/2t}\sum_{n=0}^\infty  n \sin \frac{n\pi \theta_0}{\alpha}I_{n\pi/\alpha}(\frac{\mu r_0}{t})dtd\mu $$
\end{lem}

The first term was computed in \cite[p.698]{p}, the last two follow from our previous computations.
\begin{theo}

The fair price of a CDS, taking into account the counterparty risk, is given by:
$$D_{tot}-F$$
where $F$ and $D_{tot}$ are given by the previous Lemmas.
\end{theo}

%%%%%%%%%%%%%%%%%%%%%%%%%%%%%%%%%%%%%%%%%%%%%%%%%%%%%%%%%%%%%%%
\section{First-to-default on two underlyings}
%%%%%%%%%%%%%%%%%%%%%%%%%%%%%%%%%%%%%%%%%%%%%%%%%%%%%%%%%%%%%%%

As was mentioned in the introduction, most credit derivative transactions on two default-prone instruments with time-dependent payoffs can be priced using the techniques developed in the present article. As an example, we derive the fair spread of a first-to-default contract on two underlyings. 

Most notations are as above, when we were dealing with the counterparty risk on a CDS, and we do not recall them excepted when necessary. The only difference is that, now, the two default-prone entities are treated on the same footing. We still write $C$ for the notional value of the contract and assume a given recovery rate $R$ (e.g. 40$\%$) so that, at the first default occuring before the maturity of the contract, the buyer of protection will receive $C(1-R)$ (recall this is the very definition of a first-to-default contract; on the other hand, the buyer of protection pays a continuous fee -still called the ``spread''- till the first default occurs or till the maturity of the contract if no default occurs before the maturity).

\begin{lem}\label{dl}
The default leg of the contract is given by:
$$D=C(1-R)[\int_0^T\int_0^{+\infty}e^{-rt}e^{\phi_1(\mu \cos(\alpha)-r_0\cos(\theta_0))+\phi_2(\mu \sin(\alpha)-r_0\sin(\theta_0))-{||\vec{\phi}||^2T\over 2}}\frac{\pi}{\alpha^2 t \mu }$$
$$e^{-(\mu ^2+r_0^2)/2t}\sum_{n=0}^\infty (-1)^{n+1} n \sin \frac{n\pi \theta_0}{\alpha}I_{n\pi/\alpha}(\frac{\mu r_0}{t})d\mu dt$$
$$+\int_0^T\int_0^{+\infty}e^{-rt}e^{\phi_1(\mu -r_0\cos (\theta_0))-\phi_2r_0\sin (\theta_0)-{||\vec{\phi}||^2t\over 2}}\\
\frac{\pi}{\alpha^2 t \mu }e^{-(\mu ^2+r_0^2)/2t}\sum_{n=0}^\infty  n \sin \frac{n\pi \theta_0}{\alpha}I_{n\pi/\alpha}(\frac{\mu r_0}{t})dtd\mu ]$$
\end{lem}

On the other hand, the fee leg is given by the same formula as in Lemma~\ref{fee} -this is because the coupon is served by the protection buyer till a first default occur in both cases (a CDS with counterparty risk or a first-to-default swap on two underlyings).

\begin{theo}
The fair spread $s$ of a first-to-default swap on two underlyings at inception of the contract is given by
$$s=\frac{D}{F'}$$
where $D$ is given by Lemma~\ref{dl} and $F'=Fs^{-1}$ by Lemma~\ref{fee}.
\end{theo}

%%%%%%%%%%%%%%%%%%%%%%%%%%%%%%%%%%%%%%%%%%%%%%%%%%%%%%%%%%%%%
\section{Appendix 1: Theoretical default legs for CDS}
%%%%%%%%%%%%%%%%%%%%%%%%%%%%%%%%%%%%%%%%%%%%%%%%%%%%%%%%%%%%%
We include in this section a proof of the formulas for the market price of a CDS conditional to the default of the counterparty. These are classical computations and/or variations thereof. We give therefore only short indications and refer to \cite[Chap. 3]{br} for further details.

The notations are as in Section 3; $N$ stands as usual for the cumulated standard normal distribution.
\begin{lem} On the set $\tau_2<\tau_1$, we have:
$$
P(\tau_1\leq T|{\cal F}_{\tau_2})=N(\frac{-Y_{\tau_2}-\nu_1 (T-\tau_2)}{\sigma_1\sqrt{T-\tau_2}})+e^{-2\nu_1\sigma_1^{-2}Y_{\tau_2}}N(\frac{-Y_{\tau_2}+\nu_1 (T-\tau_2)}{\sigma_1\sqrt{T-\tau_2}}),$$
where $Y_t:=\ln (\frac{V_1(t)}{v_1(t)})=\ln (\frac{V_1(0)}{K_1})+\nu_1 t+\sigma_1B_1(t)$, $\nu_1:=r-k_1-\gamma_1-\frac{1}{2}\sigma_1^2$.
\end{lem}
The lemma is a direct application of Cor. 3.1.1 in \cite{br}.

\begin{lem}
For $a,b,c\in \RM $ with $b<0$ and $c^2>a$:
$$
\int_0^ye^{ax}dN(\frac{b-cx}{\sqrt{x}})=\frac{d+c}{2d}g(y)+\frac{d-c}{2d}h(y)$$
with $d=\sqrt{c^2-2a}$, $g(y):=e^{b(c-d)}N(\frac{b-dy}{\sqrt{y}})$, $h(y):=e^{b(c+d)}N(\frac{b+dy}{\sqrt{y}}).$
\end{lem}
See Lemma 3.2.1, Fla. 3.16 in \cite{br}.
\def\ssqu{\alpha}
\def\ssqd{\beta}

\begin{cor}

We have, on $\tau_1>\tau_2$:

\

$E[e^{-r(\tau_1-\tau_2)}1_{\tau_1<s}|{\cal F}_{\tau_2}]=$
$$=e^{-\mu_{\tau_2}(\ssqd -\ssqu )}N(\frac{-\mu_{\tau_2}-\ssqu (s-\tau_2)}{\sqrt{s-\tau_2}})
+e^{-\mu_{\tau_2}(\ssqd +\ssqu )} N(\frac{-\mu_{\tau_2}+\ssqu (s-\tau_2)}{\sqrt{s-\tau_2}}),$$
where $\alpha :=\sqrt{\frac{\nu_1^2}{\sigma_1^2}+2r}$, $\beta:=\sqd$, $\mu_{\tau_2}:=\frac{Y_{\tau_2}}{\sigma_1}$.

\end{cor}
 
\def\frapp{\frac{-Y_{\tau_2}-\nu_1 (s-\tau_2)}{\sigma_1\sqrt{s-\tau_2}}}
\def\framm{\frac{-Y_{\tau_2}+\nu_1 (s-\tau_2)}{\sigma_1\sqrt{s-\tau_2}}}

Indeed by the previous lemmas,
$$E[e^{-r(\tau_1-\tau_2)}1_{\tau_1<T}|{\cal F}_{\tau_2}]=
\int_{\tau_2}^Te^{-r(s-\tau_2)}dP(\tau_1\leq s|{\cal F}_{\tau_2})$$
$$=\int_{\tau_2}^Te^{-r(s-\tau_2)}[dN(\frapp )+e^{-2\nu_1\sigma_1^{-2}Y_{\tau_2}}dN(\framm )]$$
$$=\frac{\ssqu +\ssqd}{2\ssqu}e^{-\mu_{\tau_2}(\ssqd -\ssqu )}N(\frac{-\mu_{\tau_2}-\ssqu (T-\tau_2)}{\sqrt{T-\tau_2}})+\frac{\ssqu -\ssqd}{2\ssqu}e^{-\mu_{\tau_2}(\ssqd +\ssqu )}N(\frac{-\mu_{\tau_2}+\ssqu (T-\tau_2)}{\sqrt{T-\tau_2}})+$$
$$+e^{-2\nu_1\sigma_1^{-2}Y_{\tau_2}}[ 
\frac{\ssqu -\ssqd}{2\ssqu}e^{-\mu_{\tau_2}(-\ssqd -\ssqu )}N(\frac{-\mu_{\tau_2}-\ssqu (T-\tau_2)}{\sqrt{T-\tau_2}})+\frac{\ssqu +\ssqd}{2\ssqu}e^{-\mu_{\tau_2}(-\ssqd +\ssqu )}N(\frac{-\mu_{\tau_2}+\ssqu (T-\tau_2)}{\sqrt{T-\tau_2}})],$$
and the formula follows.

%%%%%%%%%%%%%%%%%%%%%%%%%%%%%%%%%%%%%%%%%%%%%%%%%%%%%%%%%%%%%
\section{Appendix 2: First hitting time in a polyhedral domain}
%%%%%%%%%%%%%%%%%%%%%%%%%%%%%%%%%%%%%%%%%%%%%%%%%%%%%%%%%%%%%
Our results rely largely on the following two-dimensional extension of a Theorem of Daniels, see \cite[Fla 3.1]{daniels}. This natural extension is important in view of applications to multidimensional Brownian processes, particularly in the setting of credit risk. It was first announced without a proof in \cite{iy} and could be easily extended to higher dimensional cases. We could not find a proof in the literature, and include therefore a short demonstration in this appendix.

Let $Z_t^{(x_0,y_0)}$ (abbreviated to $Z_t$ when no confusion can arise), a planar Brownian starting at $(x_0,y_0)$ evolving in a polyhedral domain $D$ with absorbing boundary $\partial D$. We write $\tau$ for the first hitting time of the boundary and $f(x_0,y_0,x,y,t)$ (abbreviated to $f(x,y,t)$ when no confusion can arise) for the density of the surviving process:
$$f(x,y,t)dx dy:=P(Z_t\in (dx,dy),\tau >t)$$
We look for the distribution of hitting times: $P(\tau\in dt , Z_\tau\in dp),\ dp\in\partial D$ and can assume (up to a planar rotation) that $\partial D$ is horizontal and an upper boundary for $D$ in the neighborhood of $p$, so that $dp$ is a horizontal line element: $dp =(da,b)$, with $p=(a,b)\in \partial D$.

\begin{theo}\label{app1}We have:

\begin{eqnarray}\label{resultat}
P(\tau \in dt, Z_\tau =(da,b))&=&-\frac{1}{2}\frac{\partial}{\partial b}f(a,b,t) dadt.
\end{eqnarray}

\end{theo}

Sketch of the proof : Let us write $\phi (x,y,\Delta t)dadt$ for $P(\tau \in dt, Z_\tau\in (da, b)|Z_{t-\Delta t}=(x,y))$, where in practice we will choose $\Delta t << 1$ and study the asymptotics when $\Delta t$ goes to $0$. From Daniels' theorem \cite[Fla 3.1]{daniels} which computes the first exit density for a one-dimensional BM, together with the independence of the horizontal and vertical components of a two-dimensional BM, we get, in a neighborhood of $(a,b)$:
$$\phi(x,y,\Delta t)\cong \frac{1}{{{2}}\pi \Delta t^2}\exp{-\frac{(a-x)^2}{2\Delta t}}\cdot\exp{-\frac{(b-y)^2}{2\Delta t}}\cdot (b-y)$$
Then, 
$$P(\tau\in dt, Z_\tau \in (da,b))da^{-1}dt^{-1}=\int\limits_Df(x,y,t-\Delta t)\phi(x,y,\Delta t)dxdy$$
$$\cong \int\limits_D[f(a,y,t-\Delta t)+(x-a){f}_x^{\prime}(a,y,t-\Delta t)+\frac{(x-a)^2}{2}{f}_x^{\prime\prime}(a,y,t-\Delta t)]\phi(x,y,\Delta t)dxdy$$

By symmetry, in a neighborhood of $(a,b)$ in $D$, $\phi(x,y,\Delta t)\cong\phi(2a-x,y,\Delta t)$, so that the second term of the expansion vanishes. 
Since $\partial D$ is an absorbing boundary, $f(a,b,t-\Delta t)={f}_x^{\prime}(a,b,t-\Delta t)={f}_x^{\prime\prime}(a,b,t-\Delta t)=0$ and the third term reads:
\begin{eqnarray*}
\int\limits_D{\frac{(x-a)^2}{2}f_{x}^{\prime\prime}(a,y,t-\Delta t)\phi(x,y,\Delta t)dxdy}&\cong&{{-}}\frac{1}{\Delta t}\int_{-\infty}^{+\infty}{\frac{(x-a)^2}{2}\frac{1}{\sqrt{2\pi \Delta t}}e^{-\frac{(x-a)^2}{2\Delta t}}dx}\\
&&*\int_{-\infty}^b{(b-y)^2\frac{1}{\sqrt{2\pi \Delta t}}e^{-\frac{(b-y)^2}{2\Delta t}}\frac{\partial f_x^{\prime\prime}}{\partial y}(a,b,t-\Delta t)dy}\\
&=&O({{\Delta t}})\\
\end{eqnarray*}
Similarly, the first term reads:

\begin{eqnarray*}
\int\limits_D{f(a,y,t-\Delta t)\phi(x,y,\Delta t)dxdy} & \cong & \int_{-\infty}^b{\frac{(b-y)}{\sqrt{2\pi}\Delta t^{3/2}}e^{-\frac{(b-y)^2}{2dt}}f(a,y,t-\Delta t)dy}\int^\infty_{-\infty}{\frac{1}{\sqrt{2\pi} \Delta t^{1/2}}e^{-\frac{(x-a)^2}{2\Delta t}}dx}\\
&=&\int_{-\infty}^b{\frac{(b-y)}{\sqrt{2\pi}\Delta t^{3/2}}e^{-\frac{(b-y)^2}{2\Delta t}}f(a,y,t-\Delta t)}dy\\
&=&-\int_{-\infty}^b{\frac{(b-y)^2}{\sqrt{2\pi}\Delta t^{3/2}}e^{-\frac{(b-y)^2}{2\Delta t}}f_y^\prime(a,b,t-\Delta t)}dy\\
&=&-\frac{1}{2}f_y^\prime(a,b,t-\Delta t)
\end{eqnarray*}

The Theorem follows.

%%%%%%%%%%%%%%
%BIBLIOGRAPHIE
%%%%%%%%%%%%%%


\begin{thebibliography}{99}
\bibitem{BIS1} Bank for International Settlements, OTC Derivatives: settlement procedures and counterparty risk management, Basel, Sept. 1998. http://www.bis.org
\bibitem{BCBS} Basel Committee on Banking Supervision, The Application of Basel 2 to trading activities and the treatment of double default effects, July 2005, http://www.bis.org
\bibitem{bc} F. Black and J.C. Cox, Valuing corporate securities: Some effects of bond indenture provisions. J. Finance. 31 (1976) 351-367.
\bibitem{br} T. Bielecki and M. Rutkowski, Credit risk: modeling, valuation and hedging. Springer Finance. Springer-Verlag, Berlin, 2002.
\bibitem{cj} H.S. Carslaw and J.C. Jaeger,
Conduction of heat in solids, New York, The Clarendon Press, 1988.
\bibitem{daniels} H. Daniels, Sequential tests constructed from images, Ann. Stat. 10 (1982), 394-400.
\bibitem{gm} A. Gray, G.B. Mathews,
A treatise on Bessel functions and their applications to physics,
New York, Dover Publications, 1966.
\bibitem{hkr} H. He, W. Keirstead and J. Rebholz,
Double lookbacks.
Math. Finance 8 (1998), no. 3, 201--228.
\bibitem{hull} J. Hull, Risk Management and Financial Institutions, Pearson (2007).
\bibitem{im}  K. It\^o and  H. McKean,
Diffusion processes and their sample paths,
Die Grundlehren der mathematischen Wissenschaften, Band 125, Springer-Verlag (1974).
\bibitem{iy} S. Iyengar, Hitting lines with two-dimensional Brownian motion, SIAM J. Appl. Math. 45 (6) (1985) 983-989.
\bibitem{me} R. C. Merton, On the pricing of corporate debt: The risk structure of interest rates. J. Finance. 29, (1974), 449-470.
\bibitem{p} F. Patras, A reflection principle for correlated defaults, Stoch. Processes Appl. 116 (2006) 690-698.
\bibitem{s} Ph. Sch\"onbucher, Credit derivatives pricing models, Wiley, 2003.
\bibitem{vas} O. Vasicek, Probability of Loss on a Loan Portfolio, working paper, KMV, 1987, published as: Loan Portfolio Value, Risk, Dec. 2002.
\bibitem{w} G. Watson, A treatise on the theory of Bessel functions, 2nd ed., Cambridge University Press, 1962.
\bibitem{z} C. Zhou, An analysis of default correlations and multiple defaults, Review of Financial Studies, 14 (2) (2001), 555-576.
\end{thebibliography}
\end{document}